\documentclass[10pt]{iopart}
\usepackage{epsfig}

    \newcommand{\pathnow}{}
\begin{document}

\title[Strange Hadron Resonances  and QGP 
Freeze-out]{Strange Hadron Resonances and QGP Freeze-out}

\author{ G. Torrieri$^{\rm\dag}$ 
 and J. Rafelski$^{{\rm\dag}\,{\rm\ddag}}$}

\address{\dag\ Department of Physics,
University of Arizona, Tucson, AZ 85721,
}

\address{\ddag\ CERN-Theory Division, 1211 Geneva 23, Switzerland
}

\begin{abstract}
We describe how the abundance and distribution of hyperon
resonances can be used to probe freeze-out conditions.
We demonstrate that  resonance yields allow us to measure the time scales
of chemical and thermal freeze-outs. This should permit a
direct differentiation between the
explosive sudden,  and staged adiabatic freeze-out scenarios.
\end{abstract}

\submitto{\JPG \rm Proceedings of Strange Quark Matter 2001, Frankfurt}


\vskip -10cm \ \hfill CERN-TH/2001-367 \vskip 10cm

\section{Possible hadronization scenarios}
An important aspect in the study of heavy ion collisions is to
determine the time scale governing hadron production and the duration
of the decoupling (freeze-out) process.
At present, there is no general  consensus on the freeze-out time scale,
and hence on how much observed particles ``remember'' about their primordial source.
quark-gluon plasma (QGP) signals are visible in hadronic particles, when freeze-out is explosive, 
with sudden breakup of the fireball, and little interaction between the
 newly-formed particles  \cite{Raf00,Let00}.
In principle  the newly-formed hadrons could undergo a period
of re-interaction in a hadronic gas phase.  This re-interaction phase, the time scale
of which can be hadron-specific, could significantly alter any considered
QGP signal \cite{PBM99,CERN,Grassi,PBM01}.

The study of short-lived hyperon resonances produced in heavy ion collisions, either
by direct observation  \cite{fachini,markert}, or through their effect in
observed heavy ion $m_T$  distributions \cite{WA97},   
provides a very promising way to distinguish between these scenarios, since
natural lifetime of hyperon resonances is generally shorter than the expected
duration of the rescattering hadronic gas (HG) phase.
Therefore, the resonances signal and distribution will be 
sensitive to the duration (or indeed the existence) of this HG phase.       
In particular, rescattering of resonance decay products should deplete
the observable resonance signal, since short-lived resonances are detected 
by invariant mass reconstruction.
A significant amount of rescattering would  re-thermalize 
resonance decay products, which affects the
resulting particle's $m_T$ distributions.
This paper  analyses existing resonance data, in an atempt
to distinguish between the sudden and
staged hadronization scenarios.    

In next section \ref{spectrafits} we look the spectra of 
hyperons and show that non-rethermalized resonance decays are 
much better agreeing with the observed $m_\bot$-shapes. We than
study the directly observed strange hadron resonances in section 
\ref{dirres}. We describe the possibility of resonance quenching in 
subsection \ref{resres}, and compare in a simple model of an
opaque medium with the experimental results in subsection \ref{furres}.

\section{The role of resonances in fits}\label{spectrafits}
The precise $m_T$ spectra produced by the WA97 experiment 
\cite{WA97} provided the first opportunity
to test the sudden hadronization model.   A necessary consequence 
of explosive hadronization is that
chemical freeze-out, which fixes particle abundances, and thermal 
freeze-out, which fixes particle spectra, coincide.
Hence, it should be possible to fit both normalisation 
and $m_T$ spectra for several particles,
using the same temperature, flow and chemical potentials.   
The fitted parameters should also agree, within
error bars, with the chemical fit parameters, obtained by using just particle ratios.
Our simultaneous fit to normalized WA97 data  was successful \cite{searchqgp}.   

The sample of $K_s,\Lambda,\overline{\Lambda},\Xi,\overline{\Xi},
\Omega+\overline{\Omega}$ (divided into four centrality
bins) was fitted, using a single set of thermodynamic 
parameters for each bin, to a  Cooper-Frye distribution \cite{cooperfrye1}, truncated to eliminate
unphysical emission within a spacelike freezeout hypersurface \cite{cooperfrye2} \cite{cooperfrye3}
\begin{equation}
    \label{cooperfrye}
    \frac{d N}{m_T d m_T} = V \prod_i \lambda_i \gamma_i 
 \int d \sigma_{\mu} p^{\mu} \Theta (d \sigma_{\mu} p^{\mu} ) f(T,v,m_T)\,.
\end{equation}
Here V is the fireball volume ({\it i.e.}, an experiment-specific normalisation
parameter), $\lambda_i, \gamma_i$ are the flavor
fugacities and saturation parameters \cite{Let00},
$\sigma^{\mu}$ and $p^{\mu}$ are, respectively the 4-momentum 
and the freeze-out hypersurface, and $f(T,v,m_T)$ is
the Lorentz-transformed Boltzmann distribution, 
dependent on temperature, flow and transverse mass.

In such an  analysis \cite{searchqgp}, $T$ and the chemical parameters 
 agreed, within the error bars, with earlier abundance fits \cite{Let00}, flow and
volume increased with centrality, as expected for a thermalize expanding fireball, while the
hadronization surface velocity was consistently close to the speed of light, 
in agreement with an explosive hadronization scenario.    
$\chi^2$ profiles showed a clear, statistically significant minimum in each of the
quantities fitted.   

The only systematic deviation between statistical hadronization model and data 
is present  in the $\Omega$ spectrum.  It's unusually large enhancement
and anomalous slope are also interpreted  as evidence for a staged 
freeze-out, with $\Omega$s freezing out earlier than other particles.
However, as shown in \cite{searchqgp}, it is only the low momentum 
$\Omega$s that deviate significantly
from the expected behavior in statistical hadronization.  
Such a low-momentum yield enhancement is explained by an additional QGP
specific  physical
process, such as strange diquark correlations 
\cite{Let00}, or the formations of disoriented chiral condensates \cite{DCC}.

For the purpose of the study of the time scales in statistical 
hadronization an important direct constraint is provided by the 
prominence, in these fits, of resonance decay products.
A non-negligible fraction of hyperons (more than half in case of the $\Lambda$)
within the spectrum fitted are produced not directly, but 
through decay of  produced resonances.   Yields of heavy resonances are thermally 
suppressed because of their higher mass, but are typically
enhanced by their higher spin and isospin degeneracy.
If the resonance decays with an appreciable center of mass momentum
(as is the case in all hyperon resonances, except $\Sigma^0 \rightarrow
\Lambda \gamma$), 
the distribution of particles originating from
decays differs considerably from a thermal distribution.
Therefore, the shape of $m_\bot$ spectra which includes resonance
decay products will depend on whether the decay products 
of short-lived resonances have had the
opportunity to rethermalize or not. If  the freeze-out 
is sudden, the observed spectra are a mixture 
of direct and decay shapes, and if there is a 
phase in which the particles produced during hadronization
can undergo re-interaction, the spectrum observed is thermal, but expansion 
flow deformed.
   
\begin{figure}[tb]
\centerline{
\epsfig{width=5.8cm,clip=,figure=\pathnow 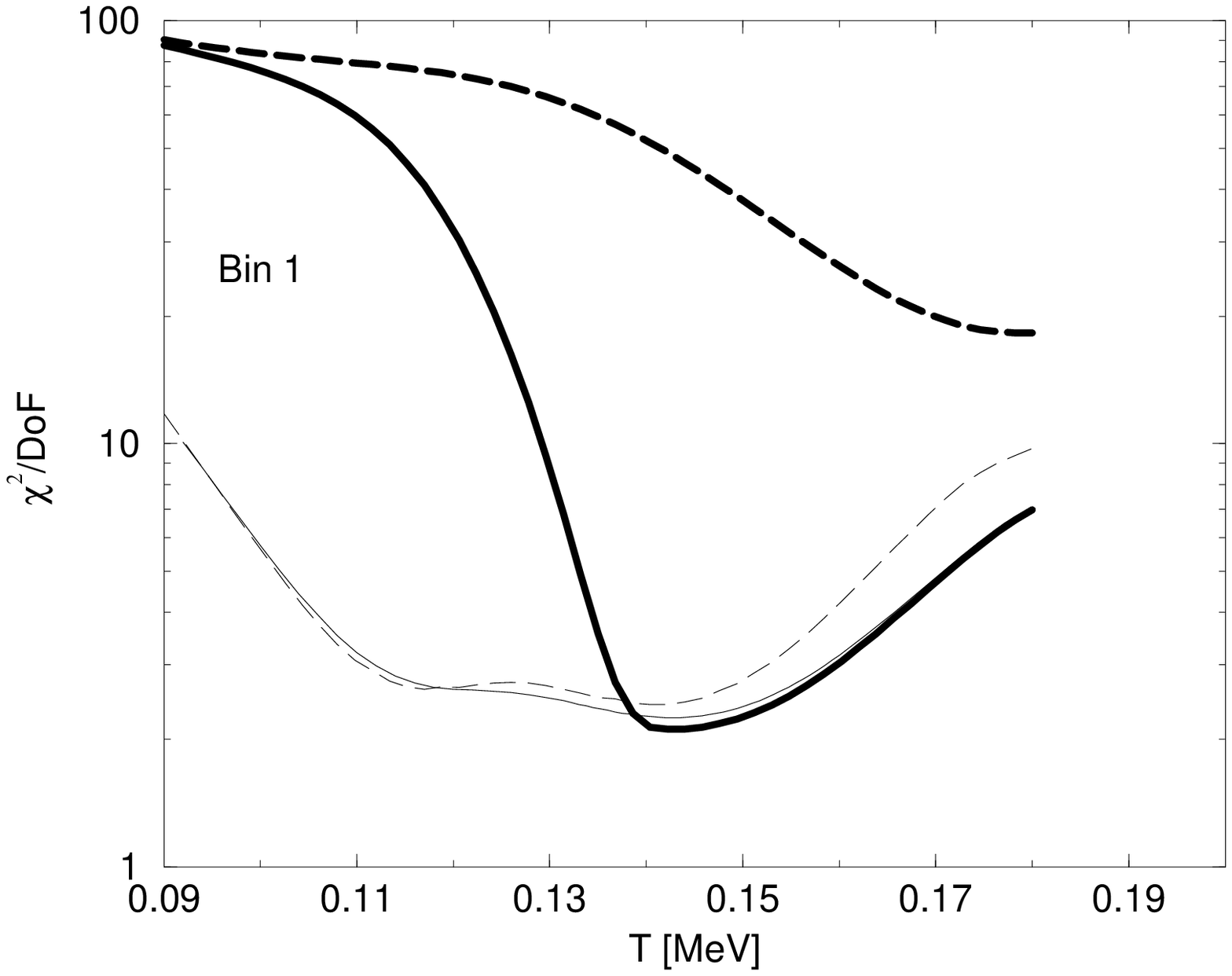}
\epsfig{width=5.8cm,clip=,figure=\pathnow 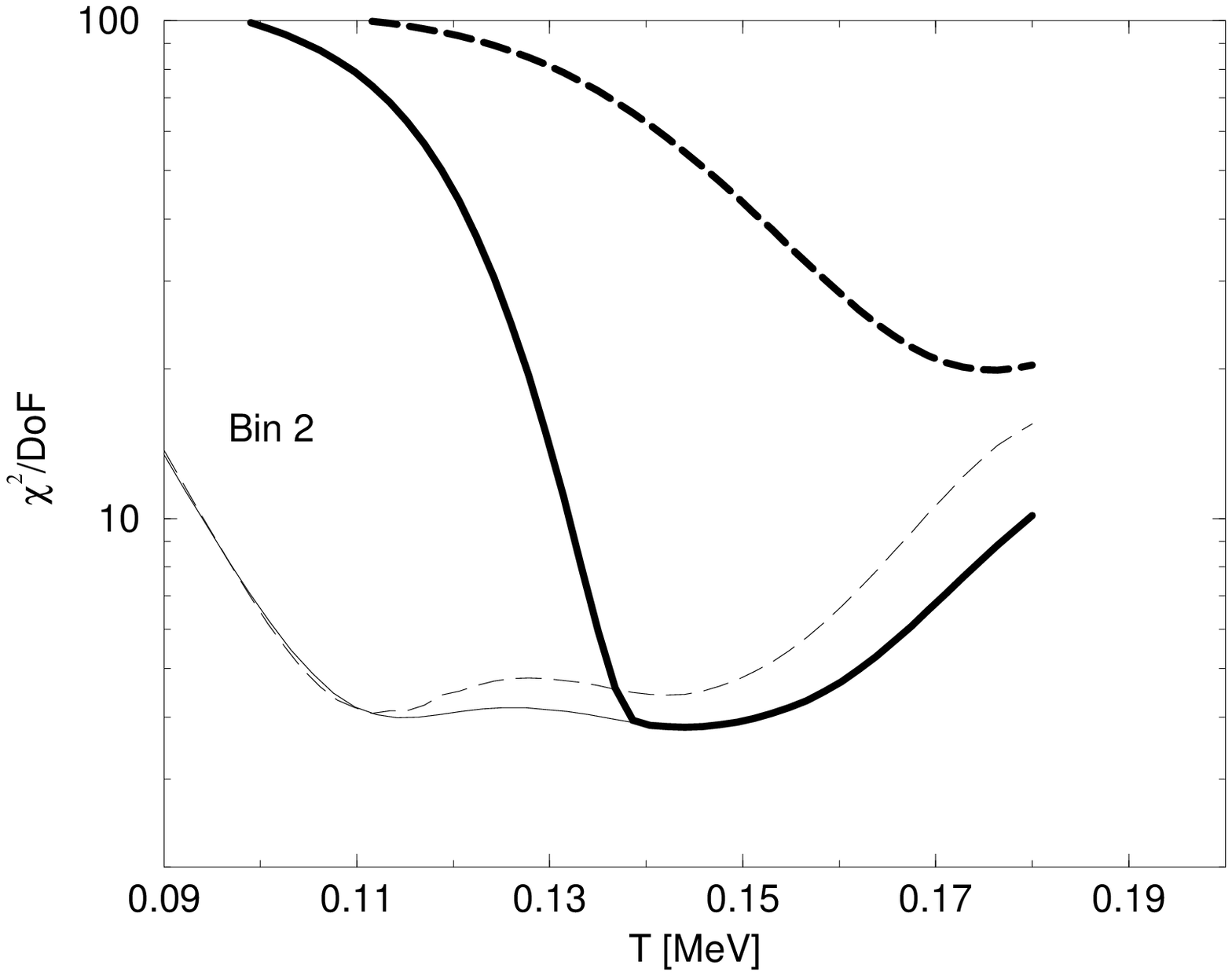}
}

\centerline{
\epsfig{width=5.8cm,clip=,figure=\pathnow 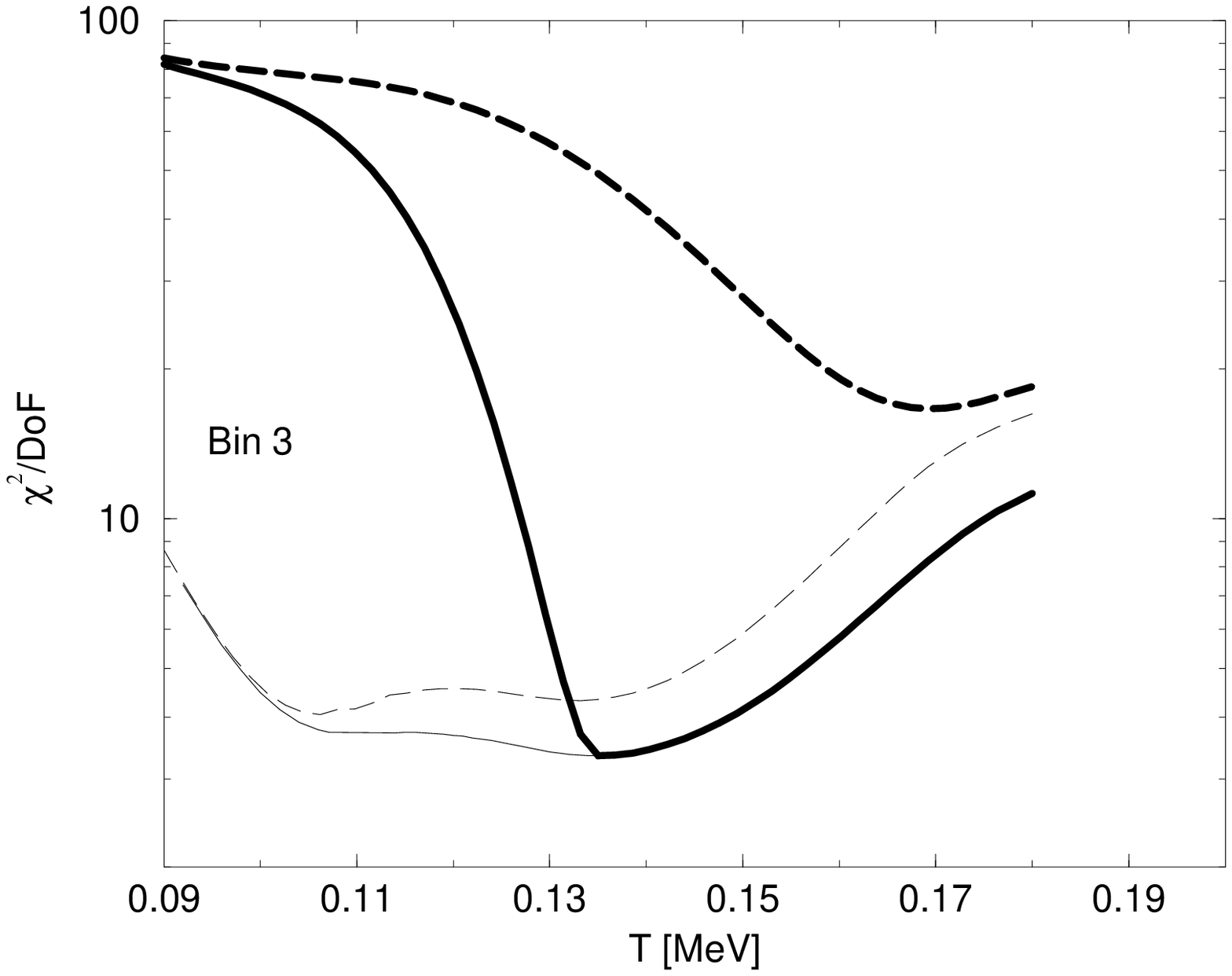}\hspace*{-0.3cm}
\epsfig{width=5.8cm,clip=,figure=\pathnow 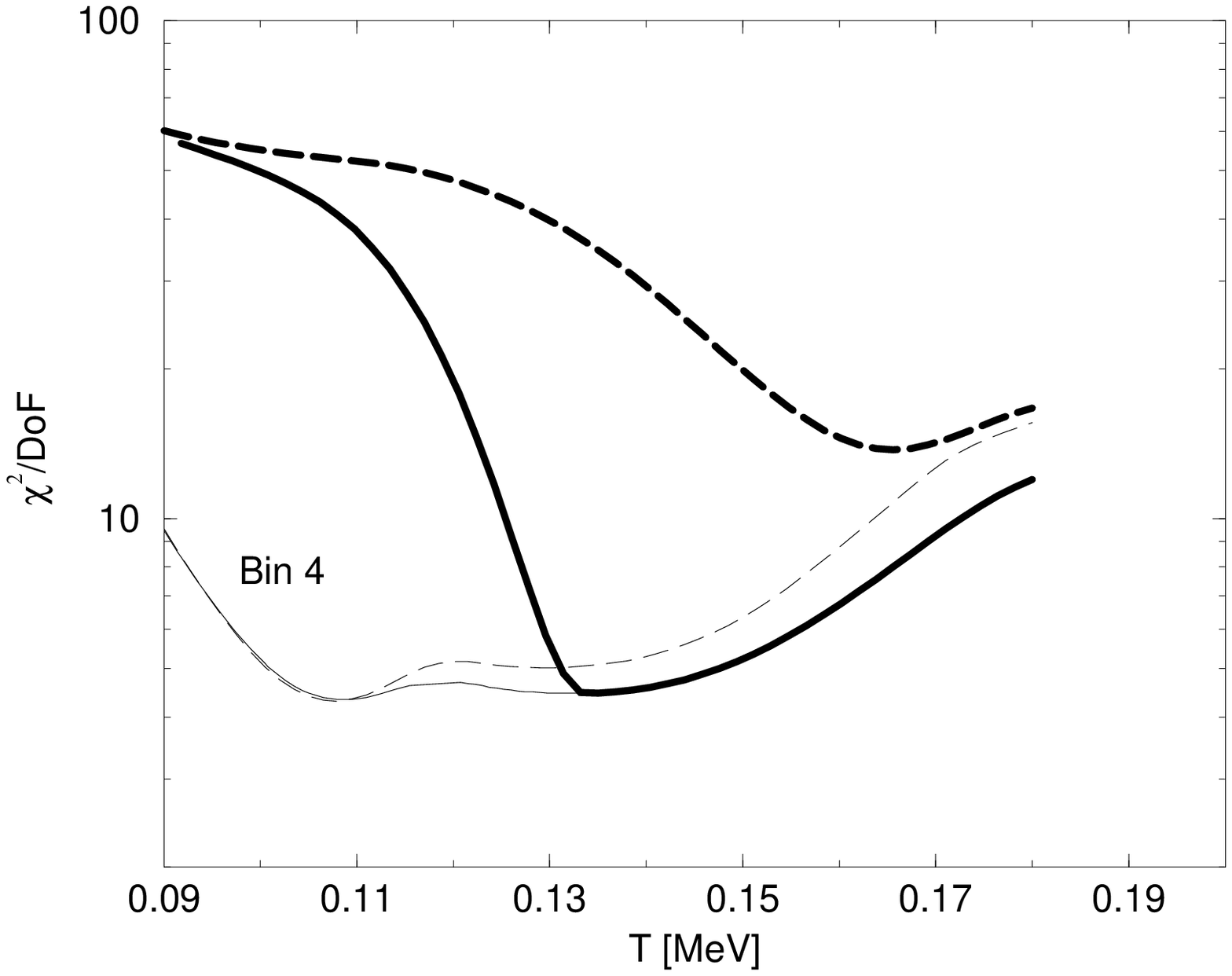}}
\caption{ 
$\chi^2$ profiles of hyperon $m_\bot$ WA97 results, 
as function of temperature $T$.
The four plots correspond to the WA97 centrality bins \protect\cite{WA97}.
Solid lines are calculated assuming that the resonance decay products
do not undergo rescattering, while dashed lines assume the decay
products rethermalize.
Thin lines show profiles calculated without enforcing
the Bose Einstein condensation limit $\gamma_q^2 \leq e^{\frac{m_{\pi}}{2T}}$,
thick lines are obtained with this constraint enforced.}
\label{chiT}
\end{figure}

In order to obtain  a final particle spectrum from a thermal distribution
of resonances, it is necessary to transform the decay
products momenta from the resonance rest frame to the fireball rest frame,
and integrate over the kinematically allowed momentum space.
This procedure has been implemented in several ways by
different authors ( \cite{resoheinz,resobecattini,resohirano} );
While numerically cumbersome, it requires no extra degrees 
of freedom within the fit.

The test for the reaction mechanism is obtained from the study of the 
$\chi^2$ distribution. A consistent lowering of $\chi^2$ once
resonances are taken into account in a hyperons $m_T$ distribution confirms that
resonance decay products emerge from the fireball unthermalized.
In contrast, if resonance decay products re-thermalize, their effect 
will be a correction to normalisation (chemical potentials
and volume) rather than slope parameters.   In such a case, 
using the decay distributions  \cite{resoheinz,resobecattini,resohirano},
will not result in a better fit to the data.

Fig.\,\ref{chiT} shows the temperature $\chi^2$ profiles (for the four
WA97 centrality bins, with bin 4 being most central), calculated
using the prescription in \cite{resoheinz}  to obtain decay spectra 
(solid lines) and 
assuming that resonance decay products rethermalize (dashed lines).
These profiles are computed by minimising 
every parameter except the temperature
(flow, hadronization surface velocity and chemical potentials).  
Thin lines are results obtained allowing unphysical range
of parameters, thick lines with the constraint on the light quark 
saturation parameter $\gamma_q$, 
corresponding to the  condensation limit of the $\pi$:
$\gamma_{\rm q}^2 \leq e^{\frac{m_{\pi}}{2 T}}$, where $m_{\pi}$ is
the pion mass. The maximum value of $\gamma_{\rm q}$ is prefered by
the hadronizing high entropy QGP phase \cite{Let00}.

We see that the condensation
constraint does not modify the value of the $\chi^2$ minimum for the solid lines 
(which correspond to non-thermalized resonance decay products), but greatly reduces
it's ambiguity, and is ruling out a freeze-out temperature lower than $\approx 145$ MeV.
As Fig.\,\ref{chiT} shows comparing thick solid with the thick dashed lines, 
the hyperon $m_\bot$ fit strongly favours a scenario 
in which the decay products of resonances
emerge from the fireball without re-interaction.
If the condensation limit on $\gamma_{\rm q}$ 
is not enforced (thin lines in \ref{chiT}), 
the $\chi^2$ is still consistently lowered
by the inclusion of unrescattered resonances. On the other hand, 
once condensation limit
is taken into account (thick lines), a model with thermalize
resonance decay products statistically fails to describe the data:
$\chi^2$ increases by a factor of 3, and the physically sensible minimum
for thermal freeze-out disappears.

We therefore conclude that the WA97 $m_\bot$ data 
strongly favours a sudden hadronization scenario 
in which the hadrons emerging from the fireball 
without much re-interaction after freeze-out.
However, a more direct way to measure the hadronization time scale
is necessary in order to firmly confirm this result.
We hope and expect that the experimental measurement of the resonance
signal will prove useful in this regard.

\section{Direct probes of sudden hadronization}
\label{dirres}
\subsection{First results on resonance yields}
\label{resres}
The resonance signal can be measured
experimentally using invariant mass reconstruction.
At present, the $K^{0*}$ has been observed by both
 NA49 \cite{friese} and STAR \cite{fachini}, and
NA49 has also measured the $\Lambda(1520)$.
It appears that while the $K^{0*}$ and $\overline{K^{*0}}$ 
are produced in the expected
amounts, the $\Lambda(1520)$ yield is severely depleted.  
In fact, so far it has proven to be
the only hyperon suppressed with respect to a proton-proton collision.

The observation of this suppression, by itself, has not been enough to settle the 
freeze-out time scale controversy. In principle such a suppression is
welcome by both camps. The proponents of sudden freeze-out argue
that freeze-out happens at  a temperature considerably lower
than that of the equilibrium QCD phase transition~\cite{Raf00}.     
The yield N of a resonance of mass $m$ at temperature $T$ obeys,
\begin{equation}
 \label{k2}
  N \propto m^2 T K_2(m/T)\propto (Tm)^{3/2}e^{m/T}\,,
\end{equation}
here $K_2$ is the modified
Bessel function. We see that  heavy resonances 
such as the $\Lambda(1520)$ are more 
suppressed, since temperature is lower. The other camp notes that
if one assumes a staged freeze-out, the suppression comes from rescattering 
of resonance decay products within HG phase.
However, in this case it is difficult to see how the $K^*$ is not
suppressed, especially considering it's larger width.

The actual situation is in our opinion  more complicated \cite{sigk0}. The 
$\Lambda(1520)$ resonance is a D-wave of the ground $\Lambda$ (Isospin=1/2) state.  
It is therefore especially susceptible to in-medium mixing 
with the lower-mass hyperons (P-wave of the
$\Sigma$,  isospin $I=3/2$ state), through reactions
such as
\begin{equation}
\label{broaden}
\pi + \Lambda(1520) \rightarrow \Sigma^* \rightarrow \Lambda\pi,\Sigma \pi\,.
\end{equation}
We therefore believe that other resonances will 
need to be measured in order to extract
freeze-out information from their abundance.  
 In particular, the non-suppression of $K^*$ s makes other
P-wave resonances good candidates for investigation.   
The $\Sigma^* (1385)$ is the least massive resonance
of this type, and for this reason, as well as it's other 
characteristics which we will explore in the
next subsection, we regard it as a promising candidate.
 
\subsection{Model and data comparison}
\label{furres}
Within 
statistical hadronization picture a particles yield 
depends on it's mass, the freeze-out temperature, and 
the chemical potentials corresponding to the
particle's quantum numbers.
If experimental acceptance does not cover the full 
$4\pi$ range, flow and hadronization
surface velocity also need to be taken into account
to determine the partial yields.

Since we are interested here primarily  in the temperature, as well 
as the evolution subsequent to hadronization,
it is indicative to pick an observable which is independent 
of as many of the above quantities as possible.
To make chemical potentials cancel out, we can examine the ratio 
of the resonance to the
``ground state'' hyperon which contains the same valence quarks.
For example, ratios such as
$\Lambda(1520)/\Lambda$,\ $\Sigma^{*}/\Lambda$, 
$K^{*0}/K^-$ and $\overline{K^{*0}}/K^+$ are
all independent of chemical potentials.  
Kaons require additional attention, because the $K^*$'s lifetime is too
short for oscillation, but the $K_s$ does oscillate.   
Therefore, the observed $K^{*0}$ and the $K^{0}$ do
not have the same valence quark composition.
We also found that such ratios, and more  generally,
ratios of particles with comparable mass,
are independent of flow and freeze-out surface 
geometry to a very good  approximation.

To get a more quantitative estimate of how 
temperature and rescattering combine in fixing resonance
ratios, a microscopic model that describes the rescattering of resonance decay
products in a baryon-rich medium needs to be considered \cite{afterburner}.
In such a model the probability of the resonance to
decay in a hadron-rich medium, as well as the interaction 
cross-section between the decay products and
the in-medium particles is allowed for.
The number of reconstructible resonance decay products is obtained solving,
\begin{eqnarray} \nonumber  
\frac{d N_{i}}{d t} &=&\ \Gamma N_{N^{*}} -N_{i} \Sigma_{j} \langle \sigma_{ij} \rangle 
\rho_{j0} (\frac{R_{0}}{R_{0}+v t})^3\,,\\
 \frac{d N_{N^*}}{d t} &=& -\Gamma N_{N^{*}}\,.      
 \label{model}
\end{eqnarray}
Here $N^*$ and $N_{i}$ are, respectively, the resonance
and decay product abundances, $\Gamma$ is the resonance width,
 $\rho_{j0}$ is the density of particle $j$ in the hadron gas at freeze-out 
and $\langle \sigma_{ij} \rangle$ is the thermally averaged interaction 
cross-section.
Final observable yields of particle $i$ are given by integrating this model from chemical
freeze-out (with a certain temperature) for an amount of time corresponding
to the interacting system's lifetime.

Because of the rapid decrease in hadron density within the fireball
due to flow (commonly estimated at around 0.5c), resonance ratios are
extremely sensitive to the lifetime of the interacting phase
compared to the resonances width.
The chemical freeze-out temperature is also important, since
it controls both the initial resonance abundance 
(through Eq.\,\ref{k2}) and the initial density of the hadron gas.

The model is however remarkably insensitive to it's other parameters,
such as the initial radius $R_0$ (which is also constrained by the
entropy per baryon,  $\approx 40$ in SPS experiments), and the cross-sections.
In particular, we found that an order of magnitude change in the interaction
cross-sections will result in ratios changing by 30 \%.
For this reason, we believe that, while our model is very
simplified (it neglects inelastic scattering
and resonance creation within a hadron gas, and  uses thermally
averaged cross-sections), the results  presented here are qualitatively
correct, and serve as a good indication of how the interesting quantities
may be extracted from experimental measurements.

\begin{figure}[tb]
\centerline{
\epsfig{width=8.4cm,clip=,figure=\pathnow 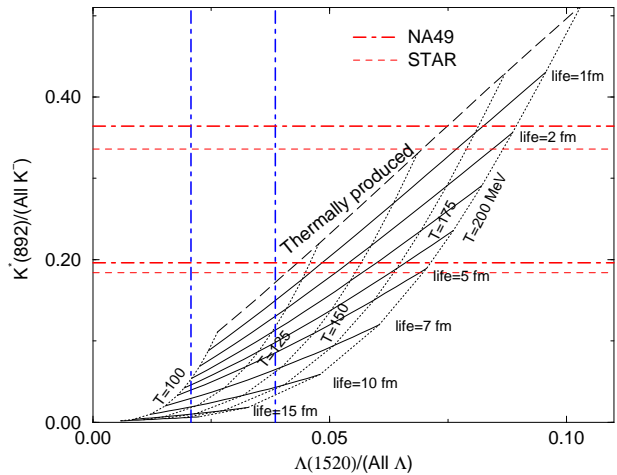}
}
\caption{ 
$K^*/K^-$ and $\Lambda(1520)/\Lambda$ ratios, shown as a function of
each other, for a range of freeze-out temperatures and lifetimes. 
The dot-dashed lines represent current experimental boundaries protect\cite{markert}.
The temperature and lifespan of freeze-out rescattering phase are shown. 
\label{kstar_lam1520}
}
\end{figure}

We now turn to discuss the available experimental results.
Fig.\,\ref{kstar_lam1520} gives the dependence 
of $K^*/K$ and $\Lambda(1520)/\Lambda$ on the chemical freeze-out temperature
and the lifetime of the interacting hadron gas phase.
It is apparent that these two quantities determine the ratios of  two
resonances of different masses and decay widths, as seen in the figure:
an experimental point on a diagram such as 
Fig.\,\ref{kstar_lam1520} is just what is needed to distinguish between
the explosive and gradual freeze-out scenarios.

\begin{figure}[tb]
\centerline{
\epsfig{width=8.4cm,clip=,figure=\pathnow 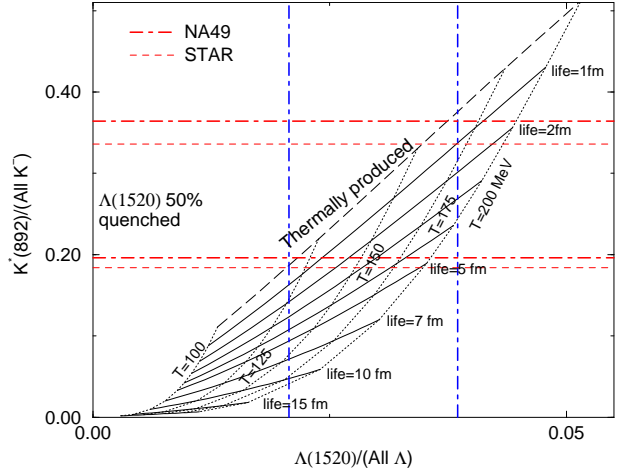}
}
\caption{ 
Same as Fig. \ref{kstar_lam1520}, assuming half of the 
$\Lambda(1520)$ are quenched, and disappear from the observed particle sample.
\label{kstarlam1520wide}
}
\end{figure}

The non-suppression of $K^*$, together with
the very strong suppression of $\Lambda(1520)$, would seem to indicate
instantaneous freeze-out with a temperature of less than 100 MeV.
However, in obtaining this result $\Lambda(1520)$ quenching was
neglected. If 50 $\%$ of the $\Lambda$ are quenched through reactions
such as those in Eq.\,(\ref{broaden}), existing data would
be perfectly compatible with the sudden freeze-out 
picture, as we show in  Fig.\,\ref{kstarlam1520wide}.

Observation of the $\Sigma^*(1385)$ yield should remove
the uncertainty introduced by the quenching phenomenon.
We have shown in \cite{sigk0}, that 
the masses and lifetimes of the $K^*$ and $\Sigma^*(1385)$ combine to constrain the
observed ratios to a very narrow yield band.
Significant deviation from this band could be a strong sign
of broadening of either of these two resonances. 
In addition, because of a coincidence between the masses of the $\Sigma^*$,
the $\Xi$ and the availability of $\Xi^*(1530)$, the explicit dependence of the
$\Sigma^*/\Xi$ ratio on temperature is very weak \cite{afterburner}.
For this reason, if fugacities and saturation parameters of the light
and strange quark are determined independently~\cite{qcdtherm}, comparison between
$\Sigma^* (1385)$, $\Xi$ and $\Lambda$ should be sufficient to disentangle
temperature and fireball lifetime conclusively.

\begin{figure}[tb]
\centerline{
\epsfig{width=8.8cm,clip=,figure=\pathnow 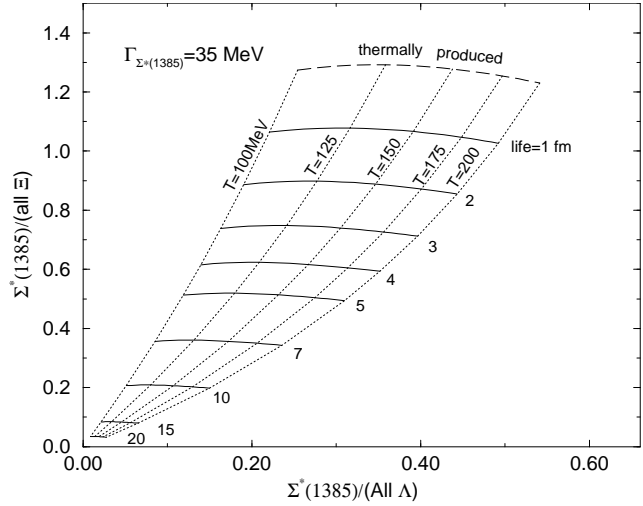}
}
\caption{ 
$\Sigma^*(1385)/\Lambda$ and $\Sigma^*(1385)/\Xi$, shown as a function of
each other for a range of freeze-out temperatures and lifetimes.
  The chemical potentials used for this figure were taken from Ref.
\protect\cite{Let00}.
\label{xi_sig1385}
}
\end{figure}

We use these properties in Fig.\,\ref{xi_sig1385},
where we show how the $\Sigma^*/\Lambda$ and $\Sigma^*/\Xi$ 
ratios depend on the freeze-out temperature and lifetime. 
We see that lines corresponding to a given  freeze-out 
temperature depend weakly on the details of the rescattering model.
This makes the $\Sigma^*/\Xi$ ratio a probe of the hadronization
temperature.
A severely depleted $\Sigma^*/\Xi$ ratio, in the absence of evidence
of broadening, would unambiguously signal a staged freeze out
in which resonance decay products have ample opportunity to rescatter.
The $\Sigma^*$ measurement could therefore
distinguish between a scenario based on the
explosive freeze-out from a super cooled QGP, 
and a high T chemical freeze-out,
followed by a reinteraction period.

\section{Conclusions}
We have shown considering the statistical significance of 
different approaches to fit the $m_\bot$ spectra of WA97 experiment \cite{WA97},
that the strategy which does not allow for resonance products to 
rescatter is greatly favoured. To find this, we needed to 
limit the magnitude of the chemical quark-pair non-equilibrium to
the maximum value compatible with the high entropy content, which is hadronizing.
The $\chi^2$ minima are very pronounced and thus offer strong evidence for 
sudden mechanism of hadronization. This finding strongly supports the 
picture of suddenly hadronizing QGP as the source of these particles.

A more direct study of hadronization lifespan is  derived from the 
observation of hyperon resonance yields. We have shown that 
combining results of several resonances allows to derive directly
information on freeze-out temperature and duration. We have 
also argued that  P-wave resonances are more suitable, than others, for this 
exercise, as they are less susceptible to in-medium modifications. 
Measurement of the $\Sigma^*(1385)$ yield, and it's
ratio to hyperons such as the $\Lambda$ and the $\Xi$, has the potential
of being just what is needed  to unambiguously extract the freeze-out 
time scale and temperature from experimental heavy ion data.

\subsection*{Acknowledgements}
We thank Ch. Markert and L. Csernai for valuable comments. 
Work supported in part by a grant from the U.S. Department of
Energy,  DE-FG03-95ER40937.


\end{document}